\documentclass[twocolumn,showpacs,aps,prl]{revtex4}

\usepackage{epsfig}

\begin{document}

\title{Theory of Supersolids beyond Mean-Field} 
\author{A. Stoffel and M. Gul\'{a}csi} 
\affiliation{
Max-Planck-Institute for the Physics of Complex Systems, 
D-01187 Dresden, Germany \\
Nonlinear Physics Centre, Australian National University,
Canberra, ACT 0200, Australia}

\date{\today}

\begin{abstract}
We investigate the newly discovered supersolid phase
by solving in random phase approximation the anisotropic 
Heisenberg model of the hard-core boson ${}^4$He lattice. 
We include nearest and next-nearest neighbor interactions 
and calculate exactly all pair correlation functions in
a cumulant expansion scheme. We describe the properties  
of the normal solid and supersolid phases and  argue that 
based on analogies to the fermionic half filled extended
Hubbard model the supersolid state corresponds to a 
bond-ordered-wave.

\end{abstract}

\maketitle

Owing to its theoretical and experimental importance, the observation \cite{KC}
of non-classical rotational inertia associated with co-existing solid and 
superfluid (SF) phases generated a lot of experimental controversy 
\cite{newexp,oldexp} and theoretical debates \cite{numerics,macro,PWA}.
However, the latest measurements \cite{newexp} clearly show that there is
an actual supersolid (SS) phase. Ref. \cite{newexp} poses 
serious limitations for previous theories \cite{numerics,macro}
but simultaneously open the door for systematic theoretical approaches, 
which is the goal of the present Letter. 

Since the work of Penrose and Onsager \cite{PO} it has been known that the SS has to 
exhibit simultaneously two types of order, namely diagonal long-range order (DLRO)
associated with the periodic modulation in a crystal and the off-diagonal 
long-range order (ODLRO) associated with the phase order in the condensate. 
Well-known for fermionic systems such as {\sl{excitonic insulator}}s \cite{review},
but, for bosons the difficult reconciliation of such puzzling
behavior has been debated since the late 1960's \cite{one,two}. 
After Kim and Chan's \cite{KC} discovery however, only two
kinds of theoretical approaches have been pursued, namely numerical
simulations \cite{numerics} and phenomenological descriptions 
\cite{macro,PWA}, while many-body theories, following the pioneering works of
Matsuda and Tsuneto; Liu and Fisher \cite{two},
have been completely neglected. 

In this Letter we fill this gap, presenting a theory beyond 
the mean-field of SS. We use a cumulant expansion and solve 
the RPA equations for the Green's functions, taking into account the
pair correlations rigorously. We will clarify the controversy over the role
of the vacancies and defects, which have long been proposed to have
a crucial role in the formation of a SS phase. We find that vacancies
and interstitials are present even at zero temperature in the SS
phase, both condense and SS may be regarded as a bond-ordered-wave 
as it exhibits alternating strength of the expectation value of the
kinetic energy term on bonds. Also, our model confirms that the SF
to SS transition is triggered by a collapsing roton minimum;
however, the SS phase is stable against spontaneously induced superflow. 

\begin{figure}[t]
\includegraphics[width=7.8cm]{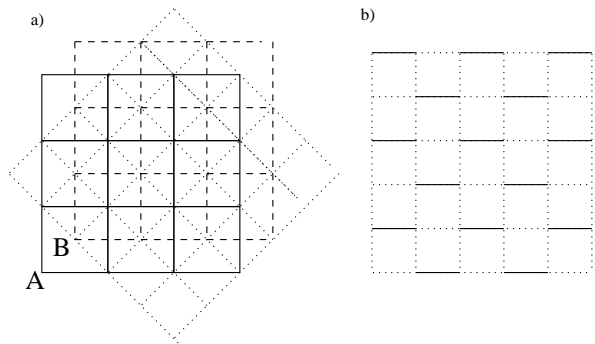} 
\caption{\label{fig:one} {\bf{a)}} A bcc lattice consists of two interpenetrating 
sc sub-lattices, $A$ (continuos line) and $B$ (long-dashed line). For simplicity
we only drawn the two dimensional case. In order to connect our result to the 
known bond-ordered-wave structures, we reduced the bcc lattice into a sc lattice
(dotted line) by unfolding the Brillouin zone. 
{\bf b)} The obtained bond-order-wave in the [100] direction is shown. 
The continuous (dashed) lines the calculated expectation
values are higher (lower).}
\end{figure}

The quantum lattice gas (QLG) model as given by Matsubara and Matsuda 
\cite{matsuda} is $K_{} = H_{QLG}-\mu N$, with
\begin{equation}
K_{} = \mu \sum_i n_i+
\sum_{i,j}u_{ij}(a_i^{\dagger}-a_j^{\dagger})
(a_i-a_j)
+\sum_{i,j} V_{ij} n_i n_j \; , 
\label{sh}
\end{equation}
where the indices $i$ and $j$ run over all lattice sites, $u_{ij}$ are the nearest 
and next nearest neighbor hopping parameters and $V_{ij}$ takes nearest and
next nearest two particle interactions into account. 
The creation and annihilation operators $a^{\dagger}_i$ 
and $a_{i}$ represent hard core bosons, i.e., $^{4}$He-atoms. Originally, 
Eq. \ref{sh} was used for SF \cite{matsuda} and later \cite{two} 
to study the possible coexistence of DLRO and ODLRO for SS. We start from a
bcc lattice, which we separate into two sub-lattices,
see, Fig. \ref{fig:one} {\bf{a)}}, establishing a natural way to define DLRO of solids: 
sub-lattice $A$ represents the ${}^4$He ions, while sub-lattice $B$
the interstitial centers. A normal solid (NS) then is given by a fully 
occupied sub-lattice $A$, while in a liquid phase the occupation number on both 
$A$ and $B$ sub-lattices are equal. 

Conventional techniques of quantum field theory are not applicable to the 
QLG model due to the lack of Wick's theorem for hard-core bosons.
Therefore we  take advantage of the well known  1-to-1 correspondence with 
spin-1/2 models imposed by the following identities:
$a^{\dagger}_j=S^x_j-i S^y_j$, $a_j=S^x_j+i S^y_j$ and 
$n_j = 1/2- S^z_j$. 
This mapping is exact and transforms the QLG model to an anisotropic 
Heisenberg model (AHM) in an external field:
\begin{equation}
H = h^z \sum_i S^z_i+\sum_{i,j}J^{\|}_{ij}S^z_iS^z_j+
\sum_{\alpha = x,y} \sum_{i,j}J^{\top}_{ij} S^\alpha_i S^\alpha_j\; , 
\label{AHM}
\end{equation}
where $J^{\|}_{ij}=V_{ij}$, $J^{\top}_{ij}=-2 u_{ij}$, 
$h^z=-\mu+\sum_{j}J^{\top}_{ij} -\sum_{j}J^{\|}_{ij}$. 
These $J$'s have to be specific for ${}^4$He. 
The interactions between the ${}^4$He atoms are controlled
by van der Waals forces and their repulsive nature at very short distances
determines negative nearest neighbor interaction $J^{\|}_1$, evoking AF
ordering in the spin language. The corresponding 
Lennard-Jones potential is short ranged and therefore it is sufficient \cite{two}
to only consider nearest and next nearest neighbor interactions. Hence, 
$J^{\|}_1=-q_1 J^{\|}_{i\in A j\in B} $, $J^{\|}_2=-q_2 J^{\|}_{i\in A j\in A}$, 
$J^{\top}_1=-q_1 J^{\top}_{i\in A j\in B}$ and 
$J^{\top}_2=-q_2 J^{\top}_{i\in A j\in A}$
where $q_1=6$ and $q_2=8$ are the number of nearest and
next nearest neighbors on the bipartite bcc lattice. 
For ${}^4$He we will use \cite{two}, $J^{\top}_1 = 1.4K$, $J^{\top}_2 = 0.5K$, 
$J^{\|}_1 = -3.8K$ and $J^{\|}_2 = -1.7K$. 
The results do not change if the $J^{\|}$'s are within $\pm 2$ 
range of these values and $J^{\|}_{B} > J^{\|}_{A}$, 
$J^{\top}$'s values remain positive. 
Different values of the $J$'s will lead more/less robust SS
phase, i.e., the SS phase will occupy a larger/smaller area in the
global phase diagram of ${}^4$He. 

The ferromagnetic (FE) and anti-ferromagnetic (AF) phases do not exhibit any off-diagonal 
long range order: they correspond to the normal fluid and NS
states of QLG, respectively. The remaining two phases of AHM are the canted
ferromagnetic (CFE) and canted anti-ferromagnetic (CAF) states.  
As the spins in the canted phases are tilted there exists 
a non-vanishing transversal component, i.e., $\langle a_i\rangle$ is non-zero. 
This is the usual order parameter for SF; hence CFE corresponds
to SF and CAF corresponds to the SS phase, respectively.  

The Heisenberg model has been studied thoroughly although there exist 
only classical mean-field solutions for the CFE and CAF \cite{two}. 
We solve the equation of motion in RPA to calculate the 
time-temperature-dependent Green's function
$G^{xy}_{ij}(t)_{Ret/Adv}=\mp i\theta(\pm t)\langle[S^x_i(t),S^y_j]\rangle ]$, 
for all four phases. In order to preserve pair 
correlations as accurately as possible, we have chosen a cumulant decoupling 
scheme to approximate higher order Green's functions. This is the 
first time that a thorough cumulant RPA many-body calculation is solved for 
the QLG and/or AHM model and as such it represents a vital step in the 
understanding of these models and their phenomena. 

The cumulant decoupling is based on the assumption that the last term of the equality
$\langle \hat{A}\hat{B}\hat{C} \rangle$ =
$\langle \hat{A}\rangle\langle \hat{B}\hat{C} \rangle$ +
$\langle \hat{B}\rangle\langle \hat{A}\hat{C} \rangle$ +
$\langle \hat{C}\rangle\langle \hat{A}\hat{B} \rangle$ -2
$\langle \hat{A}\rangle\langle \hat{B}\rangle\langle\hat{C} \rangle$+
$\langle (\hat{A}-\langle \hat{A} \rangle) 
(\hat{B}-\langle \hat{B} \rangle)
(\hat{C}-\langle \hat{C} \rangle) \rangle$
is small and thus negligible \cite{cumulant}. 
By using this we calculate exactly all the pair correlations. 
This decoupling scheme couples six Green's functions, one for each spin component 
in $x$, $y$ and $z$ direction on the two sub-lattices respectively, to a set of 
six equations. Due to the enormous number of terms (1024 in total) within these 
Green's functions we do not intend to reproduce their exact form. 
It can be shown or alternatively argued that the Goldstone theorem of gapless 
modes imposes an additional condition \cite{condition} on the mean fields in 
the SS (and SF) phase, reducing the number of order-parameters 
by two. As this condition does not apply to the NS phase their 
Green's functions are structural different.  

In the next step we calculate the phase 
diagram of AHM and derive all thermodynamic quantities, e.g., the pressure, 
the entropy and the specific heat; which are of particular interest.
Note that, by fixing the $J$'s, the chemical potential, $\mu$, remains the only variable
in the model, which we use as a fitting parameter. This is a natural choice as 
$\mu$ is related to the pressure; they are inversely proportional. In order
to derive a rigorous relationship between these two parameters we note that
$\Theta_{QLG}-\mu N=\Theta_{AHM}$, where $\Theta$ is an arbitrary potential, e.g.,
the Gibb's free energy. Then the Maxwell equation gives the desired 
relationship $(\partial p / \partial \mu)_{T,V}$ =
$(\partial N / \partial V)_{T,\mu}$ = $N_{0} (1-\epsilon) / V$,
where $N$ and $N_{0}$ denotes the number of particles and lattice sites,
respectively. Here we take into account explicitly the 
net vacancy density, $\epsilon$, as it has been long proposed that vacancies and 
defects may play a crucial role in the formation of the SS phase. 

The net density of vacancies i.e. the number of vacancies minus 
the number of interstitials:
$\varepsilon = 1 - n_A - n_B = \langle S^z_A \rangle + \langle S^z_B \rangle$,
when taken as the measure of incommensuration  
has been recently proposed \cite{PWA} as the key parameter to characterize
quantum crystals. Existing numerical calculations could only follow individual
atoms \cite{numerics} and as such found very difficult to pick out the
effects of the density of vacancies. But this is not the case in our 
many-body RPA approach. In this Letter we investigate for the first time the
effects of $\varepsilon$ on quantum solids in general and SSs 
in particular. 

\begin{figure}[t]
\includegraphics[width=6.5cm]{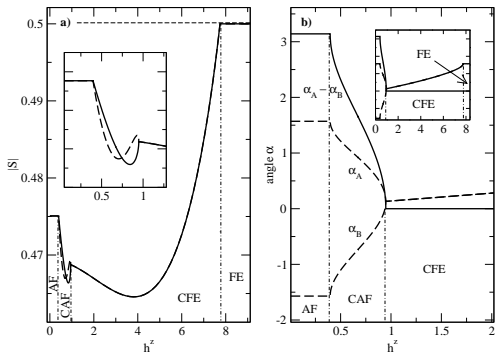} 
\caption{\label{fig:three}
The magnitude of the total spin on site $A$ and $B$ and the relative angle
$\alpha_{A/B} = \tan^{-1} \langle S^z_{A/B} \rangle / \langle S^z_{A/B} \rangle$
is plotted at $T = 0$ in {\bf a)} and {\bf b)}, respectively. In {\bf a)} 
$\vert S_A \vert$ and $\vert S_B \vert$ are identical except for the SS (canted
anti-ferromagnteic) phase. In {\bf b)} the individual $\alpha$'s are plotted
(long dashed line) and their differences (continuous line).  
}
\end{figure}

In order to understand the physical origin of the different 
phases of the QLG model, we study first the quantum fluctuation 
at $T = 0$. In  previous approaches \cite{two} 
such an analysis was impossible to perform as 
the total spin magnitude always is locked in,
or restricted to the vicinity of total spin $1/2$.
However, in our case the short and long-range correlations are
taken into account by the cumulant expansion; we can 
explicitly see the quantum fluctuations at $T = 0$. 
In Fig. \ref{fig:three} we plotted the magnitude 
of the total spin and the angle of the spin relative 
to the $x$ - $y$ - plane as a function of the external 
magnetic field $h^z$. At large $h^z$ ($\mu$, or small pressure) 
the AHM model is locked into a 
ferromagnetic phase; ${}^4$He is a normal liquid.  
There is no competition between $S^z_A$ and $S^z_B$ 
and the magnitude of the total spin is always $1/2$, i.e., 
a fully polarized state appears for any $h^z$. 

As $h^z$ ($\mu$) is lowered (increasing the pressure), 
the contribution of the transverse
term $S^x_i S^x_j + S^y_i S^y_j$, where $i, j = A, B$, 
increases until $\langle S^x_{A/B} \rangle$
and $\langle S^y_{A/B} \rangle$ become non-zero, 
giving rise to ODLRO. Consequently
$\langle S^z_{A/B} \rangle \ne 1/2$. This is the
well-know $\lambda$ transition, where 
a fully polarized FE phase becomes canted. 
From Fig. \ref{fig:three} we can see that the magnitude
of the spin is a smooth function of $h^z$  while
the relative angle between the $A$ and $B$ spins is
still vanishing, i.e., there is still no competition
between the $A$ and $B$ sites. In the language of the QLG 
model, the emergence of the CFE phase 
means $n_A = n_B \ne 0$ and, being at $T = 0$, 
condense. This is the well-known SF phase and
hence we do not go into further detail but merely emphasize
that within the AHM decsription SF appears
as a competition between $h^z S^z_{A/B}$ and the tranverse
coupling, $S^x_{A/B} S^x_{A/B} + S^y_{A/B} S^y_{A/B}$. 

As we further lower $h^z$ ($\mu$), i.e., as we increase the pressure, 
$h^z S^z_{A/B}$ starts to compete with 
the Ising coupling, $S^z_{A/B} S^z_{A/B}$ within
the already well developed ODLRO. This competition 
leads to the CAF (SS) phase, where  
$J^{\|}$ dominates over the effects of the external
magnetic field. From Fig. \ref{fig:three} we can see that
the magnitude and relative angle of the $A$ and $B$ spins differ. 
The relative angle variation gives the CAF
phase, while the difference in their
magnitude is responsible for the non-zero net density of 
vacancies. $n_{A/B} = 1/2 - \langle S^z_{A/B} \rangle$, 
Fig. \ref{fig:three} implies that $0 \le n_A \le 1/2$ and $1/2 \le n_B \le 1$
in the SS phase. Hence, as we increase the pressure, 
there will be a density transfer from the vanancies to
the interstitial sites, i.e., $n_A \longrightarrow n_B$. 
This particle density transfer 
is induced by the increasing strength of pair correlation 
$\langle S^z_A S^z_B \rangle$ and it 
only happens in the SS phase; hence it stands as the 
key in understanding the origin of this phase.
Within the QLG model, this effect will result in a alternating
strength of the expectation value of the kinetic energy term on
the bonds. This dimer order is called a bond-order-wave (BOW) 
in the half filled fermionic extended Hubbard models \cite{ozaki}.
Calculating the expectation value of the kinetic energy of the
QLG model we plot the [100] BOW corresponding to SS, see 
Fig. \ref{fig:one} {\bf b)}. Similar BOWs appear also in 
[010] and [001] corresponding to the roton minima (see below).  
Hence, we conclude that the boson condensation phenomenon of SS 
is a BOW. 

Since the 1970's \cite{roton}, it has been suggested
that a SS transition might be triggered by a collapsing roton 
minimum. In our approach we could verify this for the first time and
found that the excitation spectrum indeed goes soft at [100] 
([010] and [001]) at the SF to SS transition. 
This is true for input parameters which satisfy
$J^{\top}_1 + J^{\top}_2 + J^{\|}_1-J^{\|}_2 < 0$. In all these cases
the spectrum is not effected in the [111] direction. There are 
situations when the roton minimum collapses at [111] instead of [100], 
nameley  for $-2 J^{\top}_2 / (J^{\top}_1 + J^{\top}_2 + J^{\|}_2) > 0$. 
However, these cases induce structural instability and refer to a
transition to a different state. 

It was also conjectured \cite{roton} recently that the SF phase
in the vicinity of the SF to SS transition is unstable 
against spontaneously induced superflow and superflow associated with 
vortices. This can be easily shown not to be true in our RPA approach. 
A net superflow is either given by a moving condensate which results in  
a gradient of the phase of the wave-function or equivalently by 
a moving environment while the condensate stays at rest. Using
the latter, we add to Eq. (\ref{sh}) the term, 
$\int d^3 x \psi^{\dagger}(x)(i \hbar \mathbf{v_n} \cdot \nabla
) \psi(x)$ which transformed into the AHM language 
gives an additional contribution $\sum_{ij} J^{\times}_{ij} 
(S^x_i S^y_j -S^x_j S^y_i)$ to Eq. (\ref{AHM}). 
Here the nearest and next-nearest
neighbor cross coupling constants  
are anti-symmetric ($J^{\times}_{ij}=-J^{\times}_{ji}$)
and are zero for directions perpendicular to the 
motion of the environment $\mathbf{v_n}$. 
Calculating the excitation spectrum it is noticed that
even thought $J^{\times}$ gives a small contribution to the
spectrum, it does not influence the roton dip that triggers the
SF to SS transition. 

\begin{figure}[t]
\includegraphics[width=7.5cm]{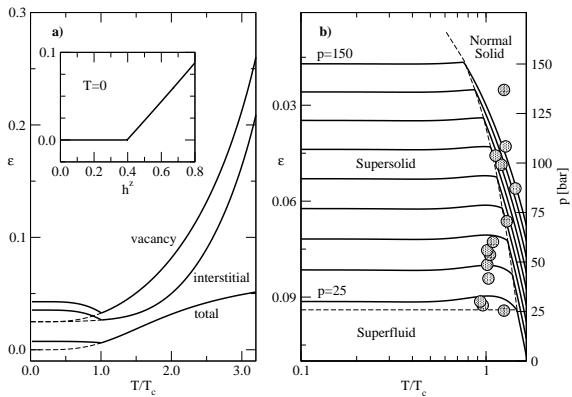} 
\caption{\label{fig:four} {\bf{a)}} Net vacancy density at the NS to 
SS transition for fixed $h^z$. 
The dashed curves are what the net vacancy density of 
the normal crystal would be without SS. Inset shows the $T =0$ limit
as a function of $h^z$. {\bf{b)}} Net vacancy density of ${}^4$He (left
axis) for different values of pressure (right axis) superimposed onto Kim 
and Chan's \cite{KC} phase diagram (filled circles).
}
\end{figure} 

At finite temperature we analyze the NS to SS transition first, as this 
is known experimentally \cite{KC}. At constant chemical potential, the 
NS to SS transition is of second order with a discontinuity in the 
net vacancy density. In Fig.\ref{fig:four} {\bf a)} we show a typical curve
of the net vacancy density at constant chemical potential. 
We have chosen a value for the chemical potenatial corresponding 
to a pressure just above the critical pressure measured by Kim and Chan 
\cite{KC}. The discontinuity in $\varepsilon$ is suggestive for a 
commensurate-incommensurate transition, when the net vacancy density 
is taken as a measure of incomensuration, as it is done in Ref. 
\cite{PWA}. 

The thin dashed curves in Fig.\ref{fig:four} {\bf a)} show what the 
net vacancy density would be if a SS phase would not appear: at 
$T \rightarrow 0$ the ground state is a perfect, commensurate 
crystal, i.e., a Mott insulator, which is not a SF. In NS the density 
of vacancies increases faster than that of interstitials with
increasing temperature, as reported in Ref.\cite{simmons}, indicating that 
thermal fluctuations favor vacancies more than interstitials. This is
not the case for the SS phase, where all three $\varepsilon$'s have
the same mean-field SF form: $\varepsilon_0 (1 - {\sqrt{\Delta / k_B T}} ) 
\exp(- \Delta / k_B T)$. Here, $\varepsilon_0 \equiv \varepsilon (T=0)$, 
shown in the inset of Fig.\ref{fig:one} {\bf{a)}}, shows that the 
SS phase is driven by the chemical potential, 
similarly to a correlation driven Mott transition: 
$\varepsilon_0 \propto (\mu - \mu_c)$, where the proportionality constant 
for our model is $( \langle S^x_A \rangle / \langle S^x_B \rangle )$. 

Recalculating the NS to SS transition as a function of the pressure
and superimposing it onto the known phase diagram \cite{KC}, 
we obtained Fig.\ref{fig:one} {\bf{b)}}. It shows 
a very good fit to the experimental points. For these isobar curves, 
as it can be seen, the net vacancy density is constant in the SS phase. 
This is because the thermal expansion of the SS phase is evanescently small.

In conclusion, we have shown that the transition into a SS
phase is of the incommensurate-commensurate type with properties 
suggesting a BOW. The AHM model in RPA does
explain most of available experimental data of the SS ${}^4$He
phase, which we will present in a future publication. Due to 
the interest in the recent experiments of Aoki, et al., \cite{newexp}, 
we briefly outline the solution to Ref.\cite{newexp} based on Eq. \ref{AHM}. 
As we have shown, the SS of ${}^4$He is a true solid hence, 
the underlying lattice is the reference frame. 
In torsional oscillator measurements, however, the reference frame is 
accelerated. Hence, the kinetic energy will acquire a new cross-coupling 
term, as presented earlier: 
$J^{\times} a \sum_{i,j} (S^x_i S^y_j - S^x_j S^y_i)$,
where $a$ is the acceleration. This will contribute with   
$J^{\top} \sum_{i,j} \cos (\phi_i - \phi_j + J^{\times} /J^{\top} a)$
to the vortex dynamics.
This implies that the motion of a single vortex/antivortex is given by
$d \bf{r_v} / dt \propto \pm \mathbf{\Omega} \times d \mathbf{v_n} / dt$, 
rather than the well-know SF equation
$d \bf{r_v} / dt \propto \pm \mathbf{\Omega} \times \mathbf{v_n}$, 
where $\mathbf{v_n}=\mathbf{r}\times\mathbf{\Omega}$ is the velocity of 
the rotating cylinder $\Omega=\Omega_0 \cos(\omega t)$. 
Thus, for a SS the signal $\rho_s / \rho$ is independent
of the frequency as seen in the Aoki, et al., experiments \cite{newexp}, i.e., 
the acceleration and not the rim velocity is the defining parameter in the 
motion of vortices.

\end{document}